\documentclass[12pt]{article}%
\usepackage{amsmath,latexsym}
\usepackage{graphicx}
\usepackage{amsmath}
\usepackage{amsfonts}
\usepackage{amssymb}%
\date{}
\begin{document}

\title{{ The compatibility of thin-shell wormholes with
    quantum field theory}}
   \author{
\large Peter K.F. Kuhfittig \footnote{kuhfitti@msoe.edu}
    \\ \\
 \small Department of Mathematics, Milwaukee School of
Engineering,
Milwaukee, Wisconsin 53202-3109, USA\\
 }
 \maketitle

\begin{abstract}\noindent
It is shown in this paper that thin-shell wormholes,
mathematically constructed by the standard cut-and-paste
technique, can, under fairly general conditions, be
compatible with quantum field theory.

\phantom{a}
\noindent
PAC numbers: 04.20.Jb, 04.20.Gz
\end{abstract}

\section{Introduction}\noindent
Wormholes are handles or tunnels in the geometry of spacetime
connecting different regions of our Universe or of different
universes.  The pioneer work of Morris and Thorne \cite
{MT88} has shown that macroscopic wormholes may not only
exist but may actually be traversable by humanoid travelers.
Wormholes can only be held open by the use of ``exotic" matter.
Such matter violates the weak energy condition.

While wormholes are predictions of Einstein's theory, quantum
field theory places severe restrictions on the existence of
traversable wormholes \cite{FR05a, FR05b, FR95, FR96}.  In
fact, Ford and Roman \cite{FR95, FR96} have shown that the
wormholes discussed in Ref. \cite{MT88} could not exist on
a macroscopic scale.  The wormholes in Refs. \cite{MSY88}
and \cite{pK99} could in principle exist, but they are
subject to extreme fine-tuning.  Fine-tuning is also required
in Refs. \cite{pK09} and \cite{pK08}, where a suitable
extension of the quantum inequalities of Ford and Roman is
used to show that Morris-Thorne wormholes can be compatible
with quantum field theory by striking a balance between the
size of the exotic region and the concomitant fine-tuning
of the metric coefficients.  Ref. \cite{pK10} continues
the theme by showing that a relatively small modification
of the charged wormhole model of Kim and Lee
\cite{KL01, KL98} can make such a wormhole compatible
with quantum field theory.

A powerful method for describing or mathematically
constructing a type of spherically symmetric wormholes from
black-hole spacetimes was proposed by Visser \cite{mV95}.
Known as \emph{thin-shell wormholes}, they are constructed
by the so-called cut-and-paste technique since their
construction requires grafting two black-hole spacetimes
together.  The thin shell is actually the junction surface.

It is proposed in this paper that under fairly general
conditions a thin-shell wormhole can be compatible with
quantum field theory.

\section{{\textbf{  Traversable wormholes}} }
   \label{S:traversable}

The spacetime geometry of a static, spherically
symmetric Lorentzian wormhole can be described by the
metric (using units in which $c=G=1$)
\begin{equation}\label{E:line1}
ds^{2}=-e^{2\Phi(r)}dt^{2}+\frac{dr^{2}}{1-b(r)/r}+r^{2}
 (d\theta^{2}+\text{sin}^{2}\theta\, d\phi^{2}),
\end{equation}
where $\Phi(r)$ and $b(r)$ have continuous derivatives.
It is usually assumed that $\Phi(r)\rightarrow 0$ and
$b(r)/r\rightarrow 0$ as $r\rightarrow\infty$, i.e., the
spacetime is asymptotically flat.  (This condition will
be automatically satisfied due to some later conditions.)
The function $\Phi(r)$ is called the \emph{redshift
function}, which must be everywhere finite to prevent an
event horizon.  We are also going to require that
$\Phi'(r)>0$.  The function $b(r)$ is called the
\emph{shape function} since it determines the spatial
shape of the wormhole when viewed, for example, in an
embedding diagram.  The minimum radius $r=r_0$ is called
the \emph{throat} of the wormhole, where $b(r_0)=r_0$.
Finally, $b'(r_0)\le 1$, referred to as the
\emph{flare-out condition} in Ref. \cite{MT88}.

Next, let us recall that the Einstein field equations
$G_{\hat{\alpha}\hat{\beta}}=8\pi T_{\hat{\alpha}
\hat{\beta}}$ imply that in the orthonormal frame, the
components of the stress-energy tensor are proportional
to the components of the Einstein tensor.  The only
nonzero components are $T_{\hat{t}\hat{t}}=\rho$,
$T_{\hat{r}\hat{r}}=p$, and $T_{\hat{\theta}\hat{\theta}}
=T_{\hat{\phi}\hat{\phi}}=p_r$, where $\rho$ is the energy
density, $p$ the radial pressure, and $p_r$ the lateral
pressure.

To hold a wormhole open, the weak energy condition (WEC)
must be violated.  The WEC states that the stress-energy
tensor $T_{\hat{\alpha}\hat{\beta}}\mu^{\hat{\alpha}}
\mu^{\hat{\beta}}$ must obey
\begin{equation}
  T_{\hat{\alpha}\hat{\beta}}\mu^{\hat{\alpha}}
  \mu^{\hat{\beta}}\ge 0
\end{equation}
for all time-like vectors and, by continuity, all null
vectors.  Using the radial outgoing null vector
$\mu^{\hat{\alpha}}=(1,1,0,0)$, the condition becomes
$T_{\hat{t}\hat{t}}+T_{\hat{r}\hat{r}}=\rho+p\ge 0$.  So if
the WEC is violated, then $\rho+p<0$.  Matter that violates
the WEC is referred to as ``exotic."

From the Einstein field equations, the components of the
stress-energy tensor turn out to be \cite{MT88, mV95}
\begin{equation}\label{E:E1}
  \rho=\frac{1}{8\pi}\frac{b'}{r^2},
\end{equation}
\begin{equation}\label{E:E2}
  p=\frac{1}{8\pi}\left[-\frac{b}{r^3}+2\left(1-\frac{b}{r}
    \right)\frac{\Phi'}{r}\right],
\end{equation}
\begin{equation}\label{E:E3}
  p_t=\frac{1}{8\pi}\left(1-\frac{b}{r}\right)
     \left[\phi''+\Phi'\left(\Phi'+\frac{1}{r}\right)\right]
     -\frac{1}{2}\left(\frac{b}{r}\right)'
            \left(\Phi'+\frac{1}{r}\right).
\end{equation}

Of particular interest in this paper is the
\emph{thin-shell wormhole} first proposed by Visser
\cite{mV95}.  The theoretical construction is also referred
to as the cut-and-paste technique and is now considered
standard.  While it is possible to start with two copies of
a generic geometry, Visser initially confined the
construction to Schwarzschild spacetimes.  In similar manner,
we will take two copies of an asymptotically flat black-hole
spacetime with (outer) event horizon $r=r_h$ and remove from
each the four-dimensional region
\[
   \Omega^{\pm}=\{r\le a|a>r_h\}.
\]
The construction proceeds by identifying (in the sense of
topology) the time-like hypersurfaces
\[
  \partial\Omega^{\pm}=\{r=a|a>r_h\}.
\]
The resulting manifold is geodesically complete and consists
of two asymptotically flat regions connected by a throat,
thereby forming a traversable Lorentzian wormhole.

The reason for our interest in thin-shell wormholes is the
following key property: all the exotic matter is confined to
an infinitely thin shell.  The property is, of course, an
idealization: we assume that the shell is extremely thin
compared to the radius of the throat.

\section{{\textbf{The quantum inequalities}} }

In this section we briefly discuss the quantum inequalities
due to Ford and Roman \cite{FR96}, slightly extended in
\cite{pK09, pK08}.  These inequalities constrain the
magnitude and time duration of negative energy and, as a
consequence, place severe restrictions on Lorentzian
wormholes.  The starting point is an inertial Minkowski
spacetime without boundaries.  So if $u^{\nu}$ is the
tangent vector to a timelike geodesic, then
$\langle T_{\mu\nu}u^{\mu}u^{\nu}\rangle$ is the expectation
value of the local energy density in the observer's frame of
reference.  So if $\tau$ is the observer's proper time and
$\tau_0$ the duration of the sampling time, then

\begin{equation}\label{E:QI1}
   \frac{\tau_0}{\pi}\int^{\infty}_{-\infty}
   \frac{\langle T_{\mu\nu}u^{\mu}u^{\nu}\rangle d\tau}
    {\tau^2+\tau_0^2}\ge -\frac{3}{32\pi^2\tau_0^4}.
\end{equation}
According to Ref. \cite{FR96}, the energy density is sampled
in a time interval of duration $\tau_0$ which is centered
around an arbitrary point on the observer's worldline so
chosen that $\tau=0$ at this point.  The inequality is
valid in curved spacetime as long as $\tau_0$ is small
compared to the local proper radii of curvature.
Applied to spherically symmetric traversable wormholes
in Ref. \cite{MT88}, it was found that none were able to
meet this condition, i.e., the throat sizes could only
be slightly larger than Planck length.

For the purpose of discussing wormholes, more convenient
forms of the above quantum inequality can be obtained.  To
do so, we need the following length scales modeled after
the length scales in Ref. \cite{FR96}, introduced in
Ref. \cite{pK09}:
\begin{equation}\label{E:rsubm}
   r_m \equiv\text{min}\left[r,
   \left|\frac{b(r)}{b'(r)}\right|,
      \frac{1}{|\Phi'(r)|},
   \left|\frac{\Phi'(r)}{\Phi''(r)}\right|\right].
\end{equation}
The reason is that the components of the Riemann
curvature tensor can be expressed in the following form:

\begin{equation}
   R_{\hat{r}\hat{t}\hat{r}\hat{t}}=
        \left(1-\frac{b(r)}{r}\right)
        \frac{1}{\frac{\Phi'(r)}{\Phi''(r)}
             \frac{1}{\Phi'(r)}}-\frac{b(r)}{2r}
     \left(\frac{1}{\frac{1}{\Phi'(r)}
         \frac{b(r)}{b'(r)}}-\frac{1}{r\frac{1}
             {\Phi'(r)}}\right)
    +\left(1-\frac{b(r)}{r}\right)
    \frac{1}{\left(\frac{1}{\Phi'(r)}\right)^2},
\end{equation}
\begin{equation}
   R_{\hat{\theta}\hat{t}\hat{\theta}\hat{t}}
  =R_{\hat{\phi}\hat{t}\hat{\phi}\hat{t}}=
    \left(1-\frac{b(r)}{r}\right)
      \frac{1}{r\frac{1}{\Phi'(r)}},
\end{equation}
\begin{equation}
  R_{\hat{\theta}\hat{r}\hat{\theta}\hat{r}}=
  R_{\hat{\phi}\hat{r}\hat{\phi}\hat{r}}
    =\frac{b(r)}{2r}\left(\frac{1}{r\frac{b(r)}{b'(r)}}
      -\frac{1}{r^2}\right),
  \end{equation}
and
\begin{equation}
  R_{\hat{\theta}\hat{\phi}\hat{\theta}\hat{\phi}}=
      \frac{1}{r^2}\frac{b(r)}{r}.
\end{equation}
The purpose of these length scales is to obtain an
upper bound for $R_{\text{max}}$, the maximum curvature:
observe that the largest value of $(1-b(r))/r$ and of
$b(r)/r$ is unity.  So disregarding the coefficient
$1/2$, we conclude that $R_{\text{max}}\le 1/r_m^2$.
Moreover, the smallest radius of curvature $r_c$ is
\begin{equation*}
   r_c\approx \frac{1}{\sqrt{R_{\text{max}}}}\ge r_m.
\end{equation*}
On this scale the spacetime is approximately Minkowskian,
so that inequality (\ref{E:QI1}) can be applied with an
appropriate $\tau_0$.

Returning to wormholes, it is proposed in Ref. \cite{FR96}
that the static frame be replaced by a ``boosted frame,"
i.e., by Lorentz transforming to a frame of a radially
moving geodesic observer moving with velocity $v$ relative
to the static frame.  In this boosted frame the energy
density $\rho'$ may be negative, so that inequality
(\ref{E:QI1}) can be applied.  In this boosted frame
(the ``primed system")
\begin{equation}
   r'_c\approx\frac{1}{\sqrt{R'_{\text{max}}}}
     \ge\frac{r_m}{\gamma},
\end{equation}
where $\gamma=(1-v^2)^{-1/2}$.  The suggested sampling
time is
\begin{equation}
  \tau_0=\frac{fr_m}{\gamma}\ll r'_c,
\end{equation}
where $f$ is a scale factor such that $f\ll1$.  The energy
density in the boosted frame is
\begin{equation}
  \rho'=T_{\hat{0}'\hat{0}'}=\gamma^2T_{\hat{t}\hat{t}}
  +\gamma^2v^2T_{\hat{r}\hat{r}}=\gamma^2(\rho+v^2p),
\end{equation}
where $v$ is the velocity of the boosted observer.  It is
stated in Ref. \cite{FR96} that the energy density does
not change very much over the short sampling time
considered here, so that $\rho'\ge -3/(32\pi^2\tau^4_0)$ is
approximately constant.  From Eqs. (\ref{E:E1}) and
(\ref{E:E2}),
\begin{equation*}
  \rho'=\frac{\gamma^2}{8\pi r^2}\left[b'(r)-v^2\frac{b(r)}{r}
  +v^2r(2\Phi'(r))\left(1-\frac{b(r)}{r}\right)\right].
\end{equation*}
For $\rho'$ to be negative, $v$ has to be sufficiently large:
\begin{equation}\label{E:velocity}
  v^2>\frac{b'(r)}{\frac{b(r)}{r}-2r\Phi'(r)
       \left(1-\frac{b(r)}{r}\right)}.
\end{equation}
In particular, at the throat, $v^2>b'(r)$.  Given $b(r)$,
inequality (\ref{E:velocity}) places a restriction on
$\Phi'(r)$.

Next, from
\[
   \frac{3}{32\pi^2\tau^4_0}\ge -\rho'
\]
we have
\begin{equation}
    \frac{32\pi^2\tau^4_0}{3}\le\frac{8\pi r^2}{\gamma^2}
    \left[v^2\frac{b(r)}{r}-b'(r)-v^2r(2\Phi'(r))
    \left(1-\frac{b(r)}{r}\right)\right]^{-1}.
\end{equation}
Using $\tau_0=fr_m/\gamma$ and dividing both sides by $r^4$,
we have (disregarding a small coefficient)
\begin{equation}
  \frac{f^4r^4_m}{r^4\gamma^4}\le\frac{1}{r^2\gamma^2}
  \left[v^2\frac{b(r)}{r}-b'(r)-2v^2r\Phi'(r)
    \left(1-\frac{b(r)}{r}\right)\right]^{-1}.
\end{equation}
Our final step is to insert $l_p$ to obtain a
dimensionless quantity:

\begin{equation}\label{E:QI2}
   \frac{r_m}{r}\le \\
    \left(\frac{1}{v^2b(r)/r-b'(r)-2v^2r\Phi'(r)
    \left(1-b(r)/r\right)}\right)^{1/4}\\
       \frac{\sqrt{\gamma}}{f}
        \left(\frac{l_p}{r}\right)^{1/2}.
\end{equation}
This is the first version of the extended quantum inequality
expressed in terms of the boosted frame.  At the throat,
where $b(r_0)=r_0$, inequality (\ref{E:QI2}) reduces to
Eq. (95) in Ref. \cite{FR96}:
\begin{equation}\label{E:QI3}
  \frac{r_m}{r_0}\le\left(\frac{1}{v^2-b'(r_0)}\right)^{1/4}
       \frac{\sqrt{\gamma}}{f}\left(\frac{l_p}{r_0}
       \right)^{1/2}.
\end{equation}
At the throat, this inequality is trivially satisfied
whenever $b'(r_0)=1$.  It was extended in Ref. \cite{pK09}.
The purpose of the extended inequality (\ref{E:QI2}) is to
include the region around the throat, instead of just the
throat itself.  [See Ref. \cite{pK09} for details.]  Also
discussed in \cite{pK09} is a second inequality that omits
both $v^2$ and $\gamma$.  The reason is that, according to
Ref. \cite{FR05a}, the boosted frame can be replaced by a
static observer.  In this frame, $r_m$ is then replaced
by $\ell_m$, the proper distance.  The extended quantum
inequality now has the slightly more convenient form
\begin{equation}\label{E:QI3}
   \frac{\ell_m}{r}\le \\
    \left(\frac{1}{b(r)/r-b'(r)-2r\Phi'(r)
    \left(1-b(r)/r\right)}\right)^{1/4}\\
       \frac{1}{f}
        \left(\frac{l_p}{r}\right)^{1/2}.
\end{equation}

\section{{\textbf{How much exotic matter?}} }

According to Ford and Roman \cite{FR96}, the exotic matter
must be confined to an extremely thin band around the throat.
This constraint is amply satisfied for a thin-shell wormhole
where the throat surface is theoretically infinitely thin.
So we are dealing with an extremely small interval, as a
result of which $\rho+p$ is approximately constant on this
interval, i.e., $\rho+p=-\eta$ ($\eta>0$).  To calculate
the volume, we need an appropriate volume measure:
$dV=4\pi r^2dr$ or $\sqrt{g}\,dr\,d\theta\, d\phi$.  So if
the shell containing the exotic matter extends from
$r=r_0$ to $r=r_1$ (on both sides  of the throat), we get
\begin{equation}\label{E:thinband}
  2\int^{r_1}_{r_0}(-\eta)dV=2\int^{r_1}_{r_0}(-\eta)
   (4\pi r^2dr)=-\frac{8\pi\eta}{3}(r_1^3-r_0^3).
\end{equation}
This volume can be made arbitrarily small by choosing
$r_1$ sufficiently close to $r_0$.  So if the shell is
infinitely thin, then the amount of exotic matter (by
volume) becomes infinitely small.  It is noted in
Refs. \cite{FR05a, FR05b, pK09}, however, that the amount
of exotic matter cannot be arbitrarily small, but as
already discussed in Sec. \ref{S:traversable}, an infinitely
thin shell is an idealization: the amount of exotic matter
could be relatively small but not arbitrarily small.

As noted earlier, having the exotic matter confined to a
very thin band is a necessary condition for the existence
of a traversable wormhole, but the condition is not
sufficient for the sought-after compatibility with
quantum field theory.  That is the topic of the next
section, which deals with wormholes in general, not
just the thin-shell type.

\section{{\textbf{A volume integral}} }

Measuring the amount of exotic matter by means of a
\emph{volume integral} involving $b$ and $\Phi$ was first
proposed by Visser \emph{et al.} \cite{VKD03} and
continued by Nandi \emph{et al.} \cite{NZK04}.  Using
Eqs. (\ref{E:E1}) and (\ref{E:E2}), it is readily
checked that
\begin{equation}\label{E:amountexotic}
  \rho+p=\frac{1}{8\pi r}\left(1-\frac{b}{r}\right)
     \left[\text{ln}\left(\frac{e^{2\Phi}}{1-b/r}
     \right)\right]'.
\end{equation}
Using the volume measure $dV=8\pi r^2dr$ from the previous
section, Eq. (\ref{E:amountexotic}) becomes
\begin{equation}
  \rho+p=(r-b)
     \left[\text{ln}\left(\frac{e^{2\Phi}}{1-b/r}
     \right)\right]'.
\end{equation}
Integrating by parts, we then obtain for the volume
\begin{equation}\label{E:volume1}
 \left. \oint(\rho+p)dV=(r-b)\text{ln}\left(
  \frac{e^{2\Phi}}{1-b/r}\right)\right|^{\infty}_{r_0}
    -\int^{\infty}_{r_0}(1-b')\left[\text{ln}
     \left(\frac{e^{2\Phi}}{1-b/r}\right)\right]dr.
\end{equation}
Regarding the boundary term, observe that since $b(r_0)=r_0$,
\begin{multline*}
  \lim_{r \to r_0}(r-b)\text{ln}\left(e^{2\Phi}\right)-(r-b)
    \,\text{ln}\left(1-\frac{b}{r}\right)=
        0-\lim_{r \to r_0}r\,\text{ln}
    \left(1-\frac{b}{r}\right)^{1-b/r}=-r_0\,\text{ln}1=0.
\end{multline*}
It is noted in Ref. \cite{NZK04} that the boundary term
also vanishes as $r\rightarrow\infty$ provided that
$\Phi(r)\sim\mathcal{O}(r^{-2})$ and
$b(r)\sim\mathcal{O}(r^{-1})$, which can likewise be
easily confirmed.

The vanishing boundary term leaves
\begin{equation}\label{E:volume2}
    \oint(\rho+p)dV=
    -\int^{\infty}_{r_0}(1-b')\left[\text{ln}
     \left(\frac{e^{2\Phi}}{1-b/r}\right)\right]dr.
\end{equation}
It is easy to show that this integral is well behaved near
$r=r_0$.  For the upper limit, the conditions $\Phi(r)\sim
\mathcal{O}(r^{-2})$ and $b(r)\sim\mathcal{O}(r^{-1})$
carry the day.

Now recall from the previous section that the interval
$[r_0,r_1]$ containing the exotic matter can be made as
small as desired.  The only way to obtain a small value
for the integral in Eq. (\ref{E:volume2}) is by letting
$b'$ be close to unity.  For convenience, let us now
restate inequality (\ref{E:QI3}):
\begin{equation}\label{E:QI4}
   \frac{\ell_m}{r}\le \\
    \left(\frac{1}{b(r)/r-b'(r)-2r\Phi'(r)
    \left(1-b(r)/r\right)}\right)^{1/4}\\
       \frac{1}{f}
        \left(\frac{l_p}{r}\right)^{1/2}.
\end{equation}
Observe that at the throat, where $b(r_0)=r_0$, the
inequality is trivially satisfied whenever $b'(r_0)=1$.
Moving away from the throat, the denominator must still
be 0 or close to 0, calling for an appropriate adjustment
of the redshift function $\Phi(r)$.  Such an adjustment
is in principle possible for the following reason: it is
shown in Ref. \cite{pK09} that, for any of the typical
shape functions, $b(r)/r-b'(r)>0$; at the same time,
$\Phi'(r)>0$, an assumption that was brought out earlier.
(The only other assumptions made are that $\Phi(r)$
remains finite and that $\Phi(r)\sim\mathcal{O}(r^{-2})$,
which have no direct bearing on the problem.)  So
inequality (\ref{E:QI4}) can be satisfied in the
neighborhood of the throat, provided that $b'(r)$
remains close to 1 (while gradually declining).  But
the small value of the left side of
Eq. (\ref{E:volume2}) forces $b'$ to be close to 1 near
the throat. The thin-shell wormhole is therefore
compatible with quantum field theory under fairly
general conditions.

\section{The volume integral theorem}
The volume integral for determining the amount
of exotic matter has been extended to \cite{NZK04}
\begin{equation}\label{E:int1}
   \Omega=\int^{2\pi}_0\int^{\pi}_0\int^{\infty}_{r_0}
   (\rho+p)\sqrt{-g_4}\,dr\,d\theta\,d\phi.
\end{equation}
Applied to an infinitely thin shell of radius
$a$, it is sometimes assumed that
\[
\rho(r)=\delta (r-a)\sigma (a),
\]
where $\sigma(a)$ is the energy density of
the thin shell and $\delta$ is the Dirac
delta function \cite{ES05}.  Using the orthonormal
basis $e_{\hat{t}}=[f(r)]^{-1/2}e_t$,
$e_{\hat{r}}=[f(r)]^{1/2}e_r$, $e_{\hat{\theta}}
=(r^2)^{-1/2}e_{\theta}$, and $e_{\hat{\phi}}=
[r^2\text{sin}^2\theta]^{-1/2}e_{\phi}$,
we get $\sqrt{-g_4}=r^2\text{sin}\,\theta$.
Since an infinitely thin shell does not exert
any radial pressure, we now have
\begin{multline}\label{E:int2}
  \int^{2\pi}_0\int^{\pi}_0
  \int^{\infty}_{-\infty}\delta(r-a)
  \sigma(a)r^2\text{sin}\,\theta
  \,dr\,d\theta\,d\phi
  =\int^{2\pi}_0\int^{\pi}_0\sigma(a)r^2
  \left|_{r=a}\text{sin}\,\theta\,
  d\theta\,d\phi \right.\\
  =4\pi\sigma(a)a^2.
\end{multline}
For a Schwarzschild spacetime, where
\[
    \sigma(a)=-\frac{1}{2\pi a}
    \sqrt{1-\frac{2M}{a}},
\]
the result is
\[
    \Omega=-2a\sqrt{1-\frac{2M}{a}};
\]
moreover, for large $a$, $\Omega\approx
-2a$.  This can lead to a huge amount
of exotic matter since we are using
geometrized units.  For example, if
$M=0.1$ m, and $a=1$ m, then
$\Omega=-1.8$ m, which exceeds the mass
of Jupiter in absolute value even though
the shell is infinitely thin.  At the
opposite end, $\Omega$ can be made
arbitrarily small by choosing $a$
sufficiently close to the event horizon
$r=2M$.  The result is a black-hole
mimicker \cite{LZ08}, which may be
indistinguishable from a black hole
at a distance.  So while the use of
the delta function is mathematically
correct, when applied to an infinitely
thin shell, this use does not appear
to be appropriate for expressing the
density $\rho$ of the shell.  The
outcome also conflicts with the
results in the previous section.

To see the reason for the discrepancy
between the two models, let us use the
prelimit form of the delta function,
i.e.,
\begin{equation}\label{E:delta}
   \rho(r)=\sigma(a)
   \begin{cases}
    \frac{1}{\epsilon},&a\le r\le a+\epsilon\\
    0, &\text{otherwise}
    \end{cases}
\end{equation}
and then take the limit as $\epsilon
\rightarrow 0$.  Now Eq. (\ref{E:int1})
becomes
\begin{multline*}
  \Omega=\text{lim}_{\epsilon\rightarrow 0}
  \int^{2\pi}_0\int^{\pi}_0\int^{a+\epsilon}
  _a\sigma(a)\frac{1}{\epsilon}\sqrt{-g_4}\,dr
  \,d\theta\,d\phi\\=\text{lim}_{\epsilon
  \rightarrow 0}\int^{2\pi}_0\int^{\pi}_0
  \int^{a+\epsilon}_a\sigma(a)\frac{1}
  {\epsilon}r^2\text{sin}\,\theta\,
  \,drd\theta\,d\phi\\
  =\text{lim}_{\epsilon\rightarrow 0}
  4\pi\sigma(a)\frac{1}{\epsilon}\frac{1}{3}
  [(a+\epsilon)^3-a^3]
  =4\pi\sigma(a)a^2
\end{multline*}
by L'Hospital's rule, in agreement with
Eq. (\ref{E:int2}).

The use of the prelimit form now allows a
comparison between the two models: as
$\epsilon\rightarrow 0$, the density increases
beyond any bound, i.e., $\rho(r)\rightarrow
\infty$ by Eq. (\ref{E:delta}).  Since we do
not normally talk about the ``value" of the
delta function but only about the values of
integrals involving the delta function, the
assumption that $\rho(r)\rightarrow \infty$
is not inconsistent with the usual
formalism, but which is the preferred model?
It was noted earlier that
$\rho(r)$ is approximately constant on a
small interval.  So from a physical
standpoint, $\rho(r)$ may be large, but
it is definitely finite, and this implies
that as $\epsilon\rightarrow 0$, the area
under the graph goes to zero, so that
$\Omega\rightarrow 0$, as in the previous
section.  By contrast, the use of the
delta function leads to the unjustifiable
assumption that the area remains fixed at
$\sigma(a)\cdot1$ for every interval
$[a,a+\epsilon]$, no matter how small.


\begin{thebibliography}{3}

\bibitem{MT88}M.S. Morris and K.S. Thorne, Wormholes in spacetime
and their use for interstellar travel: A tool for teaching general
relativity, \emph{Am. J. Phys.} \textbf{56}, pp. 395-412, 1988.
\bibitem{FR05a}C.J. Fewster and T.A. Roman, On wormholes with
arbitrarily small quantities of exotic matter, \emph{Phys. Rev. D}
\textbf{72}, Article ID 044023 (15 pages), 2005.
\bibitem{FR05b}C.J. Fewster and T.A. Roman, Problems with wormholes which
involve arbitrarily small amounts of exotic matter, arXiv: gr-qc/0510079.
\bibitem{FR95}L.H. Ford and T.A. Roman, Averaged energy conditions and
quantum inequalities, \emph{Phys. Rev. D} \textbf{51}, pp. 4277-4286, 1995.
\bibitem{FR96}L.H. Ford and T.A. Roman, Quantum field theory constrains
traversable wormhole geometries, \emph{Phys. Rev. D} \textbf{53}, pp.
5496-5507, 1996.
\bibitem{MSY88}M.S. Morris, K.S. Thorne, and U. Yurtsever, Wormholes,
time machines, and the weak energy condition, \emph{Phys. Rev. Lett.}
\textbf{61}, pp. 1446-1449, 1988.
\bibitem{pK99}P.K.F. Kuhfittig, A wormhole with a special shape function,
\emph{Am. J. Phys.} \textbf{67}, pp. 125-126, 1999.
\bibitem{pK09}P.K.F. Kuhfittig, Theoretical construction of Morris-Thorne
wormholes compatible with quantum field theory, arXiv: 0908.4233.
\bibitem{pK08}P.K.F. Kuhfittig, Viable models of traversable wormholes
supported by small amounts of exotic matter, \emph{Int. J. Pure Appl.
Math.} \textbf{44}, pp. 467-482, 2008.
\bibitem{pK10}P.K.F. Kuhfittig, On the feasibility of charged wormholes,
\emph{Cent. Eur. J. Phys.} \textbf{9}, pp. 1144-1150, 2011.
\bibitem{KL01}S.-W. Kim and H. Lee, Exact solutions of charged wormholes,
\emph{Phys. Rev. D} \textbf{63}, Article ID 064014 (5 pages), 2001.
\bibitem{KL98}S.-W. Kim and S.P. Kim, Traversable wormhole with classical
scalar fields, \emph{Phys. Rev. D} \textbf{58}, Article ID 087703
(4 pages), 1998.
\bibitem{mV95}M. Visser, Lorentzian wormholes - from Einstein to Hawking
   (American Institute of Physics, New York, 1995).
\bibitem{VKD03}M. Visser, S. Kar, and N. Dadhich, Traversable wormholes with
arbitrarily small energy condition violations, \emph{Phys. Rev. Lett.}
\textbf{90}, Article ID 201102 (4 pages), 2003.
\bibitem{NZK04}K.K. Nandi, Y.-Z. Zhang, and K.B. Vijaya Kumar, Volume integral
theorem for exotic matter, \emph{Phys. Rev. D}  \textbf{70}, Article ID
127503 (4 pages), 2004.
\bibitem{ES05}E.F. Eiroa and C. Simeone, Thin-shell wormholes in dilaton
gravity, \emph{Phys. Rev. D} \textbf{71}, Article ID 127501 (4 pages), 2005.
\bibitem{LZ08}J.P.S. Lemos and O.B. Zaslavskii, ``Black hole mimickers:
regular versus singular behavior," Phys. Rev. D \textbf{78}, 024040, 2008.


 \end{thebibliography}
\end{document}